\documentclass[12pt]{iopart}

\usepackage{amsfonts}
\expandafter\let\csname equation*\endcsname\relax
\expandafter\let\csname endequation*\endcsname\relax
\usepackage{amsmath}
\usepackage{amssymb}
\usepackage[square,numbers,sort&compress]{natbib}
\usepackage{graphicx}
\usepackage{color}\usepackage{url}
\usepackage[bookmarks, breaklinks]{hyperref}
\usepackage{youngtab}

\begin{document}

\title[Exact density profiles and symmetry classification for  strongly interacting fermions]{Exact density profiles and symmetry classification for  strongly interacting multi-component Fermi gases in tight waveguides}

\author{Jean Decamp}
\address{Universit\'e de Nice-Sophia
  Antipolis, Institut Non Lin\'eaire de Nice, CNRS, 1361 route des Lucioles, 06560 Valbonne, France}

\author{Pacome Armagnat}
\address{Universit\'e Grenoble-Alpes, INAC-PHELIQS, F-38000 Grenoble, France}
\address{CEA, INAC-PHELIQS, F-38000 Grenoble, France}

\author{Bess Fang}
\address{LNE-SYRTE, Observatoire de Paris, PSL Research University, CNRS, Sorbonne Universit\'es, UPMC Univ. Paris 06, 61 avenue de l'Observatoire, 75014 Paris, France}

\author{Mathias Albert}
\address{Universit\'e de Nice-Sophia
  Antipolis, Institut Non Lin\'eaire de Nice, CNRS, 1361 route des Lucioles, 06560 Valbonne, France}

\author{Anna Minguzzi}
\address{Universit\'e Grenoble-Alpes, LPMMC, BP166, F-38042 Grenoble, France}
\address{CNRS, LPMMC, BP166, F-38042 Grenoble, France}

\author{Patrizia Vignolo}
\address{Universit\'e de Nice-Sophia
  Antipolis, Institut Non Lin\'eaire de Nice, CNRS, 1361 route des Lucioles, 06560 Valbonne, France}

\date{\today}

\begin{abstract}
We consider a mixture of one-dimensional strongly interacting Fermi gases with up to six components, subjected to a longitudinal harmonic confinement. In the limit of infinitely strong repulsions we provide an exact solution which generalizes the one for the two-component mixture. We show that an imbalanced mixture under harmonic confinement  displays partial spatial separation among the components, with a structure which depends on the relative population of the various components.  Furthermore, we provide a symmetry characterization of the ground and excited states of the mixture introducing and evaluating a suitable operator, namely the conjugacy class sum. We show that, even under external confinement,  the gas has a definite symmetry which corresponds to the most symmetric one  compatible with the imbalance among the components. This generalizes the predictions of the Lieb-Mattis theorem for a fermionic mixture with more than two components. 
\end{abstract}
\pacs{05.30.-d,67.85.-d,67.85.Pq}


\section*{Introduction}

Ultracold atomic gases made with rare-earth elements cooled to quantum degeneracy and subjected to two-dimensional optical lattices provide a beautiful realization of the model of  one-dimensional multicomponent Fermi gases with strong and equal intercomponent repulsion among the species \cite{Pagano2014}. 

In the absence of external harmonic confinement in the longitudinal direction,
the system of  multi-component fermions with intercomponent delta interactions is  a generalization of the Yang-Gaudin Hamiltonian \cite{McGuire,Gaudin67,Yang67} and can be solved by nested Bethe Ansatz \cite{Sutherland68}. The homogeneous system  may also be described at low energy as a multi-component Luttinger liquid. This model has been extensively studied in the context of electronic multichannel systems  \cite{Glazman1993,Giamarchi_book}  and exotic condensed matter materials \cite{Giamarchi_book,Murray2013,Orignac2012,Frahm1993,Murray2011}. 

The experiments with ultracold atoms  are characterized by the presence of an external longitudinal confinement, which can be well approximated as harmonic. In this case, the Bethe Ansatz solutions do not apply, however,
a special integrable case is provided by the limit of infinitely strong repulsions among the species. In this regime, corresponding to the Tonks-Girardeau regime for ultracold bosonic atoms, fermions belonging to different components further fermionize, as has been experimentally shown in Ref.~\cite{SelimJochim}: the wavefunction vanishes at contact and they can be mapped onto a noninteracting Fermi gas in the same external confinement with particle number corresponding to the total number of fermions in the mixture, following the same idea as the original Girardeau solution for bosons \cite{Gir1960}. An additional difficulty in the multicomponent case is that the manifold displays a large degeneracy \cite{GirMin2007}. This is associated to the arbitrariness in fixing the relative phase once two fermions belonging to different components get in contact and then exchange their mutual position. The degeneracy is however broken at finite interactions, where one expects a  unique ground state \cite{Blume13,CuiHo14}. The ground state branch can be obtained by performing a strong-coupling $1/g$ expansion, $g$ being the interaction strength among the fermions. This provides a unique way to identify in the degenerate manifold what would be the ground and excited states at finite interactions  \cite{CuiHo14,Volosniev2014}.

 In this work, we explore this type of solution for multicomponent Fermi mixtures with a number of components that ranges from two to six as in the $^{173}$Yb experiment of Ref.~\cite{Pagano2014}. We study in particular the density profiles of the imbalanced mixtures, thus generalizing the works of \cite{NJPPolishgroup,Pecak2016}. As main result, we find a complex, inhomogeneous spatial structure. We then explore the symmetry property of both the ground and excited state branches of the solution, introducing and evaluating suitable operators for the mixture, ie  the transposition  class-sum and the three-cycle class sum operators \cite{Katriel1993,Novolesky1994}. This allows us to test (and generalize) the Lieb-Mattis theorem \cite{LiebMattisPR} for the trapped multicomponent mixtures, showing that the ground state carries  the  most symmetric configuration allowed by the imbalance among the components.

\section{Model}

We consider a system of $N$ fermions of equal mass $m$, divided in $r$ species with population $N_1,N_2,\ldots,N_r$. We assume that all components are subjected to the same  harmonic potential $V(x)=m\omega^2 x^2/2$, as is the case of fermions in optical traps. The fermions belonging to different species interact with each other via the contact potential $v(x-x')=g\delta(x-x')$, where $g$ is the interaction strength, and $\delta(x)$ is the Dirac delta function. The total Hamiltonian reads
\begin{equation}
{\cal H}=\sum_{i=1}^N\left[ - \frac{\hbar^2}{2m} \frac{\partial^2}{\partial x_i^2} + \frac 1 2 m \omega^2 x_i^2\right]+ g \sum_{i<j} \delta(x_i-x_j),
\end{equation}
where $x_1,\ldots,x_N$ are the coordinates of the fermions.  The effect of contact interactions can be replaced by a cusp condition on the many-body wavefunction :
\begin{equation}
\frac{\hbar^2}{2m g}\left[\left(\frac{\partial\Psi}{\partial x_i}-\frac{\partial\Psi}{\partial x_j}\right)_{x_i-x_j\to 0^+}-\left(\frac{\partial\Psi}{\partial x_i}-\frac{\partial\Psi}{\partial x_j}\right)_{x_i-x_j\to 0^-}\right]=\Psi(x_i=x_j).
\end{equation} 
In the following,  we will focus on the impenetrable limit $g\to\infty$. In this case the cusp condition imposes that the many-body wavefunction vanishes when two particles meet, ie $\Psi(x_j=x_\ell)=0$ for each pair $\{j,\ell\}$. This condition is exactly satistisfied by the fully antisymmetric solution  $\Psi_A(x_1,\ldots,x_N)$ of  $N=N_1+N_2+...+N_r$ noninteracting fermions in the same confining potential, corresponding to a Slater determinant constructed with single-particle wavefunctions $\phi_0,\ldots,\phi_{N-1}$. To construct the exact solution, 
we further note that the behaviour of the many-body wavefunction under exchange between two fermions belonging to different components is not fixed by symmetry, and requires to be fixed by additional conditions. Hence,  we consider a general solution of the form \cite{Deuretzbacher,Volosniev2014}
\begin{equation}
\Psi(x_1,\ldots,x_N)=\sum_{P\in S_N}a_P\chi(x_{P(1)}<\cdots<  x_{P(N)})\Psi_A(x_1,\ldots,x_N),
\label{Eq3:Psi}
\end{equation}
where $\chi(x_1<\cdots<  x_N)$ is the indicator function of the sector $\{x_1<\cdots<  x_N\}\subset\mathbb{R}^N$, ie it is $1$ within the sector and $0$ everywhere else and $S_N$ the permutation group of $N$ elements. Here, both $\Psi$ and $\Psi_A$ are assumed to have unit normalization, and the choice of the coefficients $a_P$ will be detailed below. Note that we need only to determine  $S=\frac{N!}{N_1!\ldots N_r!}$  coefficients $a_P$:  the wavefunction is antisymmetric under exchange of fermions belonging to the same component, and this property is already encoded in $\Psi_A$. This observation allows us to restrict ourself to the so-called snippet basis \cite{Deuretzbacher,Fang2011}, ie consider only the global permutations modulo the permutations of particles belonging to  the same species. 

{\it General solution}

In order to determine the $a_P$ coefficients for the ground state manifold, denoted $a_1,\ldots,a_S$ in the snippet basis, we use the same method as in Ref.\cite{Volosniev2014}. It  consists in a perturbative expansion of the energy to first order in $1/g\to 0$, ie we write $E=E_A- (\hbar^4/m^2) K/ g$ where $E_A$ is the energy associated with $\Psi_A$ and  $K=- (m^2/\hbar^4) ({\partial{E}}/{\partial{g^{-1}}})$ is proportional to the interaction energy, related to the Tan's contact coefficient in the two-component case \cite{Zwe11}. 
We then use the Hellmann-Feynman theorem and the cusp condition to write 
\begin{equation}
K=\frac{1}{4\left\langle\Psi|\Psi\right\rangle}\sum_{i<j}\int\mathrm{d}x_1\ldots\mathrm{d}x_N\delta(x_i-x_j)\left[\left(\frac{\partial\Psi}{\partial x_i}-\frac{\partial\Psi}{\partial x_j}\right)_{x_i-x_j\to 0^+}-\left(\frac{\partial\Psi}{\partial x_i}-\frac{\partial\Psi}{\partial x_j}\right)_{x_i-x_j\to 0^-}\right]^2,
\end{equation} 
where we have used the natural units, namely the harmonic oscillator length $a_{ho}=\sqrt{\hbar/m\omega}$ as unit length and the harmonic oscillator energy $\hbar\omega$ as unit energy.
Separating the integral over the different sectors $\{x_{P(1)}<\cdots<x_{P(n)}\}\subset\mathbb{R}^N$ and recalling  that $\Psi_A$ is normalized to one, we finally obtain
\begin{equation}
K=\frac{\sum_{P,Q\in S_N}(a_P-a_Q)^2\alpha_{P,Q}}{\frac{1}{N!}\sum_{P\in S_N}a_P^2},
\end{equation}
where the $\alpha_{P,Q}$ coefficients are non zero if the sectors $P$ and $Q$ differ only by transposing two adjacent coordinates, and in this case we have, using permutational symmetry,
\begin{equation}
\label{alphak}
\alpha_{P,Q}=\alpha_k=\int_{x_1<\cdots<  x_N}\mathrm{d}x_1\ldots\mathrm{d}x_N\delta(x_k-x_{k+1})\left[\frac{\partial \Psi_A}{\partial x_k}\right]^2,
\end{equation}
for $k\in\{1,\ldots,N-1\}$. Intuitively, $\alpha_k$  can be seen as the energy cost of an exchange between particles of different
species at positions $k$ and $k+1$. Note that, thanks to the parity invariance of this problem, we also have $\alpha_k=\alpha_{N-k}$, so that we have only $\lfloor N/2 \rfloor$ coefficients to compute.

In order to find the wavefunction which corresponds to the ground state at finite, large interactions, the next step is to find the solutions that minimize the energy, ie that maximize $K$. To do so, we impose that $({\partial K}/{\partial a_i})=0$ for all $a_i$. This turns out to be equivalent to the  diagonalization problem 
\begin{equation}
{\mathbf V}\vec A=K\vec A,
\label{Eq:inlineVAeqKA}
\end{equation}
with $\vec A=(a_1,\ldots,a_S)^T$ and ${\mathbf V}$ is a $S \times S$ matrix depending only on the $\alpha_k$ coefficients. More precisely, the ${\mathbf V}$ matrix is defined in the snippet basis by
\begin{equation}
\label{volosniev}
V_{ij}=\left\{
\begin{array}{ll}
 -\alpha_{i,j} & \mbox{if } i\ne j \\ \sum_{k\ne i}\alpha_{i,k} & \mbox{if  } i=j\end{array}\right.,
\end{equation}
where the $\alpha$ coefficents are defined as in Eq. \eqref{alphak} (see also \cite{Deuretzbacher2014,Loft2016}).

In order to compute the $\alpha_k$ coefficients, we use the following expression for $\Psi_A$, based on a Vandermonde determinant result \cite{GirWriTri01, Forrester03}
\begin{equation}
\Psi_A(x_1,\ldots,x_N)=\frac{1}{\sqrt{N!}}C_N\prod_{k=1}^{N}e^{-x_k^2/2}\prod_{1\le i<j \le N}(x_i-x_j),
\end{equation}
where $C_N=\frac{1}{\pi ^{N/4} \prod   _{k=0}^{N-1} \sqrt{2^{k}k!}}$. We then have after some algebra
\begin{equation}
\left[\frac{\partial\Psi_A}{\partial x_k}\right]_{x_k=x_{k+1}}=\frac{1}{\sqrt{N!}}C_Ne^{-x_k^2}\prod_{\substack{i=1 \\ i\neq k,k+1 }}^Ne^{-x_i^2/2}(x_i-x_k)^2\prod_{\substack{1\le j<\ell \le N \\j,l\neq k,k+1 }}(x_j-x_{\ell}),
\end{equation}
and thus, using again the Vandermonde formula
\begin{equation}
\left[\frac{\partial\Psi_A}{\partial x_k}\right]_{x_k=x_{k+1}}^2=\frac{2^{2N-3}e^{-2x_k^2}}{\pi N!(N-1)!(N-2)!}\left[\det    \left[ \phi_{i-1}(x_j)\right]\right]^2_{N-2\times N-2}
\prod_{\substack{i=1 \\ i\neq k,k+1 }}^N(x_i-x_k)^4.
\end{equation}
Finally, permutation and parity invariances yield
\begin{equation}
\begin{split}
\alpha_k=&\frac{2^{2N-3}}{\pi N!(N-1)!(N-2)!}\frac{1}{(k-1)!(N-k-1)!}\int_{-\infty}^{+\infty}\mathrm{d}x_ke^{-2x_k^2}\sum_{P,Q\in S_{N-2}}\epsilon (P) \epsilon (Q)\\
&\prod_{\substack{i=1 \\ i\neq k,k+1}}^N\int_{L_k}^{U_k}\mathrm{d}x_i(x_i-x_k)^4\phi_{P(i)-1}(x_i)\phi_{Q(i)-1}(x_i),\\
\end{split}
\label{Eq10:alphak}
\end{equation}
where the integration limits are $(L_k,U_k)=(-\infty,x_k)$ if $i<k$ and $(x_k,+\infty)$ otherwise.
An alternative derivation of the coefficients $\alpha_k$ can be found in \cite{Deuretzbacher2014}.

\section{Density profiles}
As the first application, using Eq.(\ref{Eq3:Psi}) together with the solution of Eq.(\ref{Eq:inlineVAeqKA}) for the coefficients $a_P$ and Eq.(\ref{Eq10:alphak}) for the weigths  $\alpha_k$  we determine the exact density profile of each component of the mixture, according to the definition 
\begin{equation}
n_\nu(x)=N_\nu \int \prod_{j=1}^N dx_j \delta(x-x_\nu) \Psi^*(x_1,...x_N) \Psi(x_1,...x_N),
\end{equation}
where we have indicated by  $x_\nu$ one of the coordinates corresponding to a fermion belonging to the $\nu$-th component of the mixture. In the following, for each given mixture, we focus on the ground state and first many-body excited state with a symmetry different from the ground state one.

\begin{figure}
\begin{center}
\includegraphics[width=0.45\linewidth]{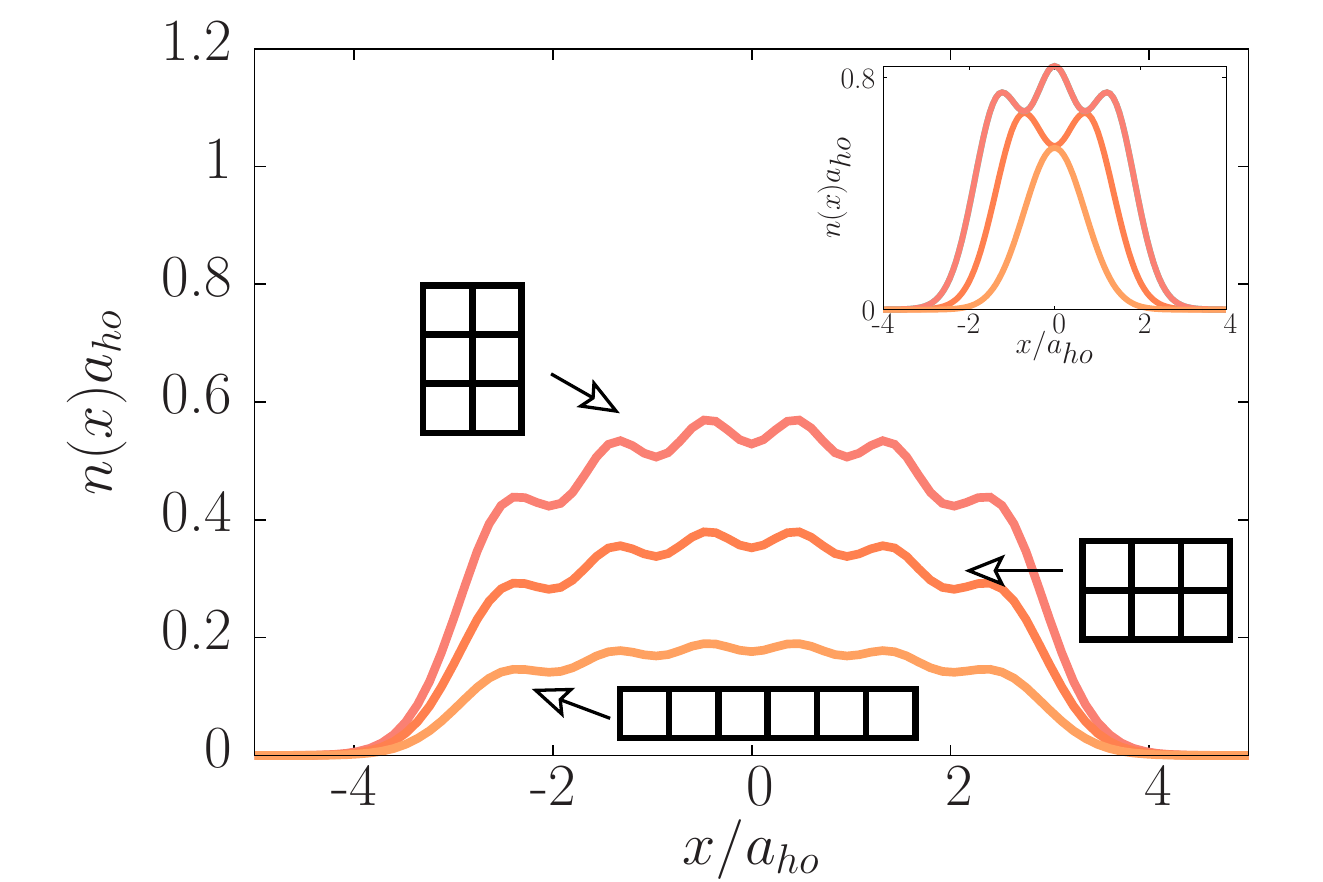}
\includegraphics[width=0.45\linewidth]{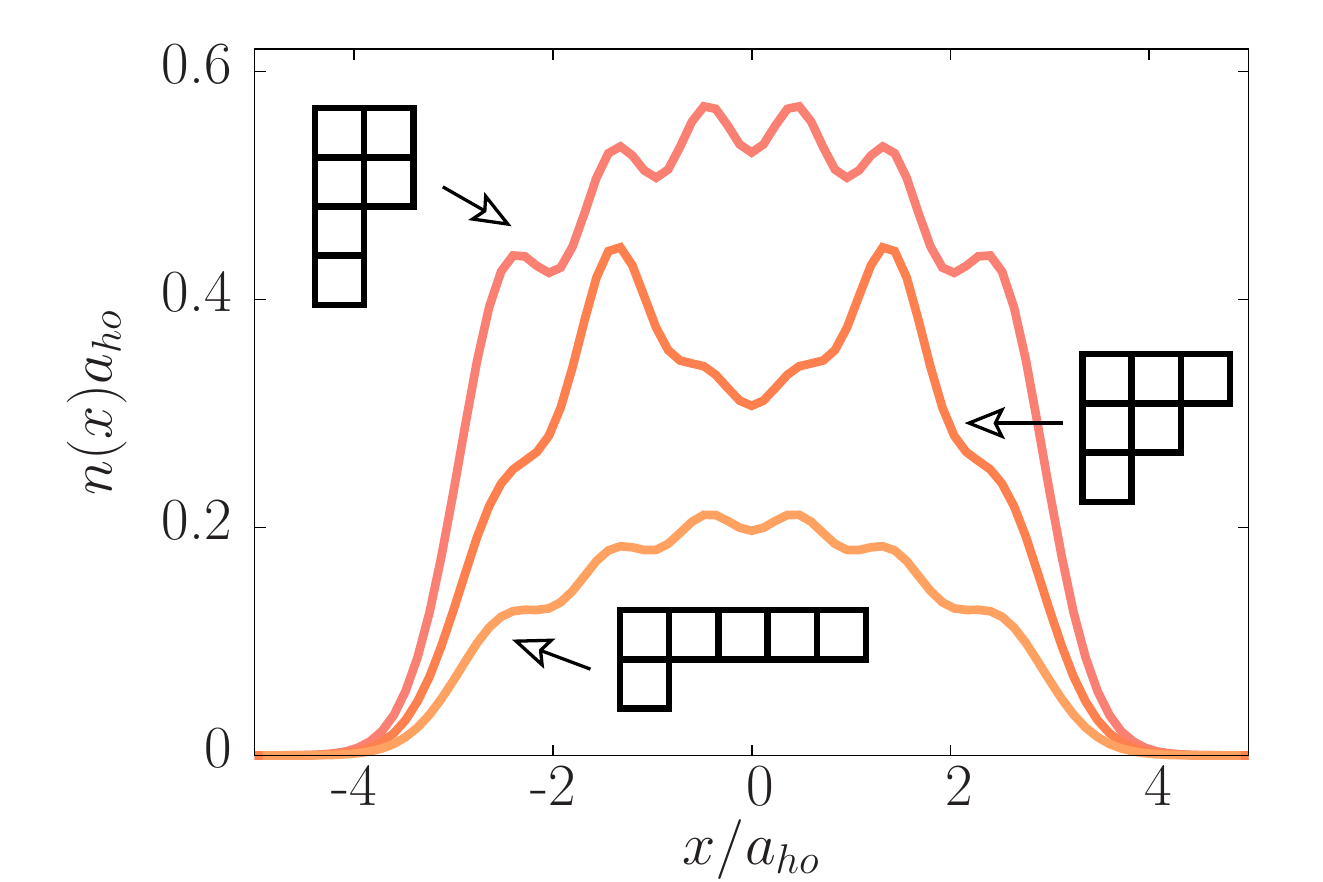}
\end{center}
\caption{\label{fig1} Density profiles for the ground state (left panel) and for the first many-body excited state with a symmetry different than the ground state (right panel) for three balanced mixtures (ie with the same number of particles in each species $N_1=\cdots=N_r$) of strongly interacting Fermi gases, with different numbers of components $r=2,3,6$ and total particle number 
 $N=6$ (from top to bottom: $N_\nu=3,2,1$).  The density profiles are the same for each component of the mixture, ie $n_1(x)=n_2(x)=...=n_r(x)$. The inset shows the corresponding ground state density profiles for the case of the corresponding mixtures of noninteracting fermions. The corresponding Young tableaux (see text) are also shown in the panels near each profile.}
\end{figure}

Quite generally, for all the ground state density profiles we find that the total density $n(x)=\sum_{\nu=1}^r n_\nu(x)$ coincides with the one of a noninteracting $N$-particle Fermi gas under external confinement, as previously reported \cite{massignanNJP}.  For equal populations in the various species , the ground state density profile is the same for all the species and coincides apart to a normalization factor with the one of a noninteracting Fermi gas with $N$ particles in harmonic confinement, which is characterized by $N$ peaks \cite{VignoloMinguzziPRL}, as shown in Fig.\ref{fig1}. The excited states display a variety of profiles, where in particular the two-component mixture does not change and the six-component one shows small deviations with respect to the corresponding ground-state density profiles, while the three-component mixture displays a different, two-peak structure. These differences may be accounted for by considering the different symmetry of the excited states in the three cases, see  Sec.~\ref{sec:sym}.

\begin{figure}
\begin{center}
\includegraphics[width=0.45\linewidth]{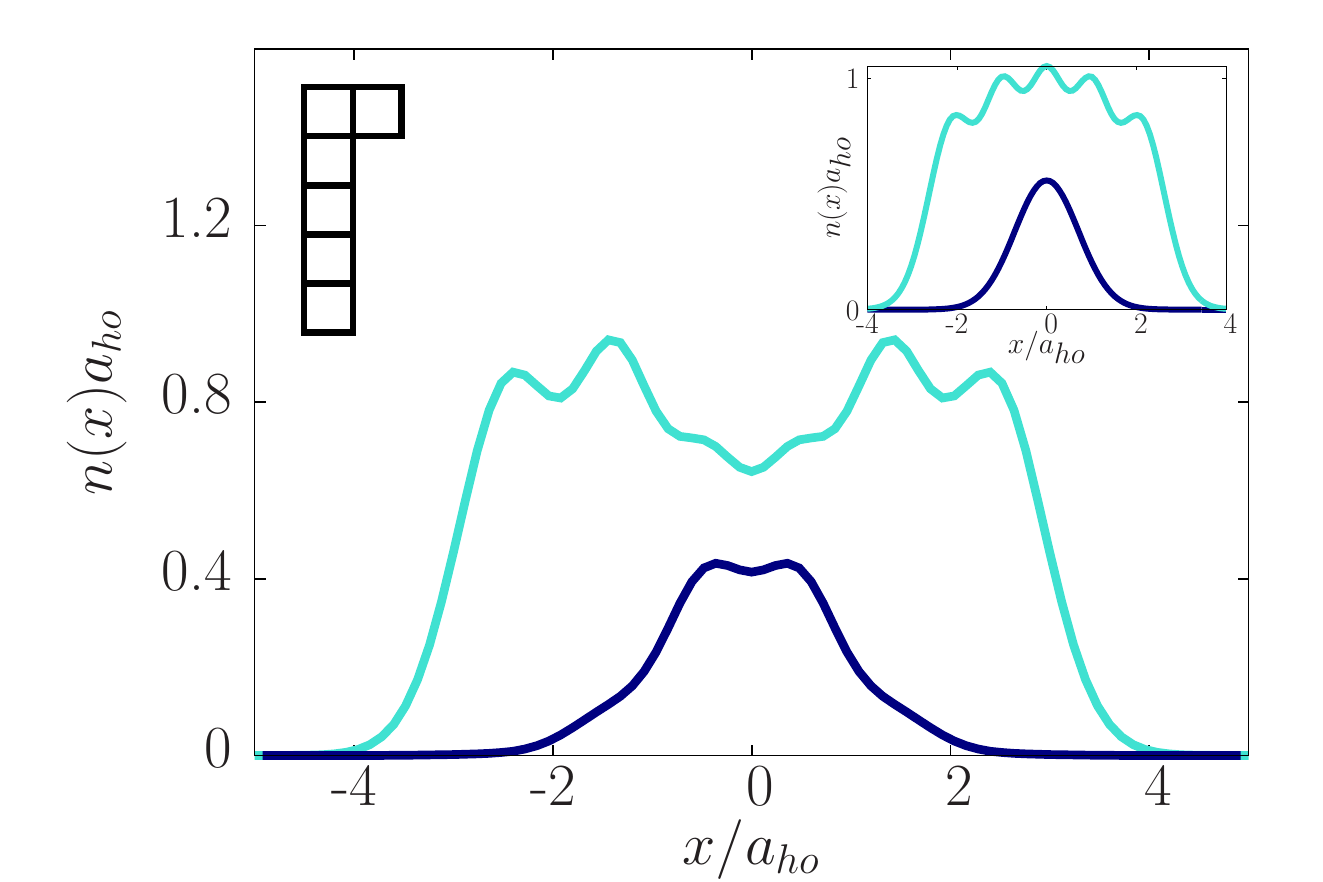}
\includegraphics[width=0.45\linewidth]{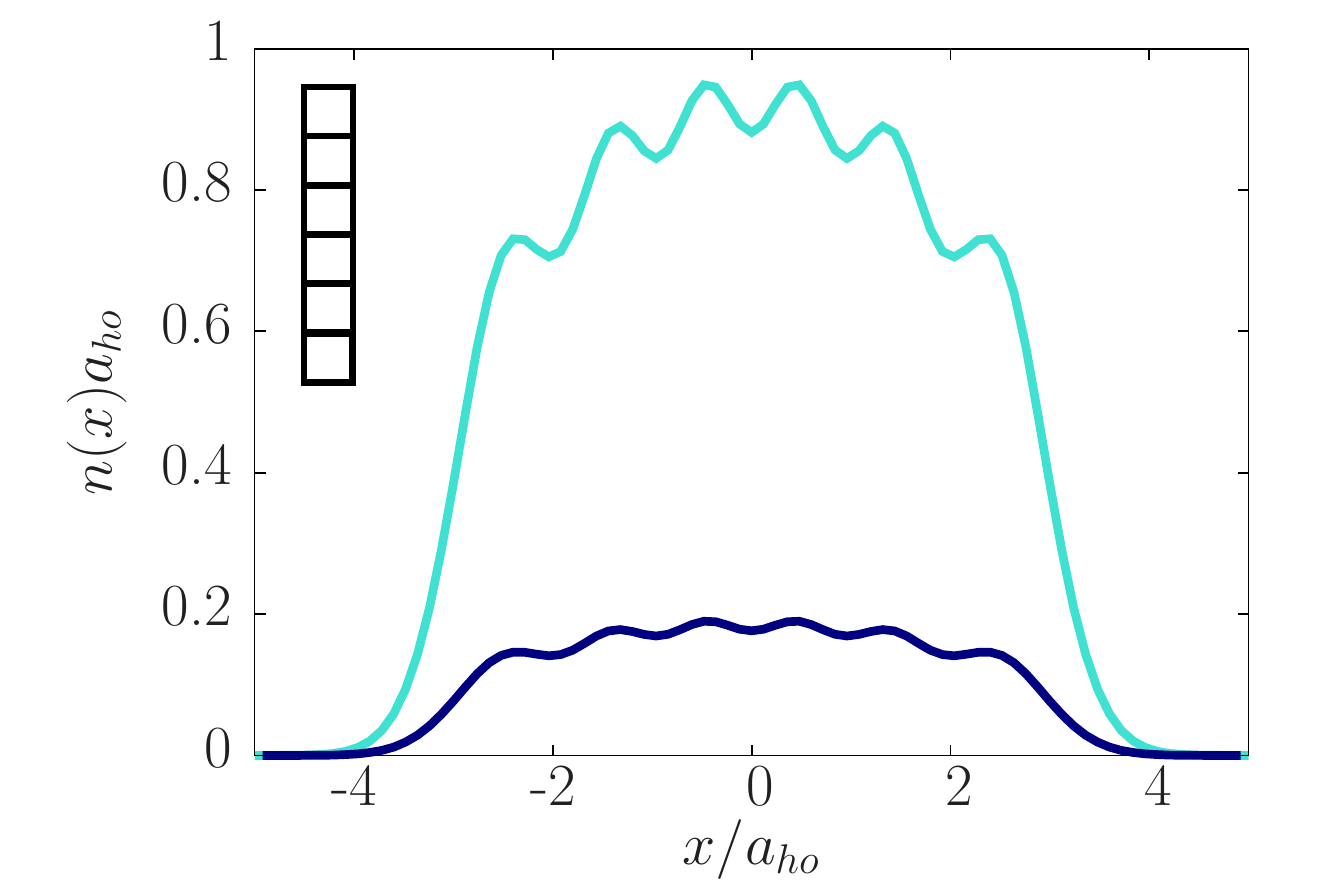}
\includegraphics[width=0.45\linewidth]{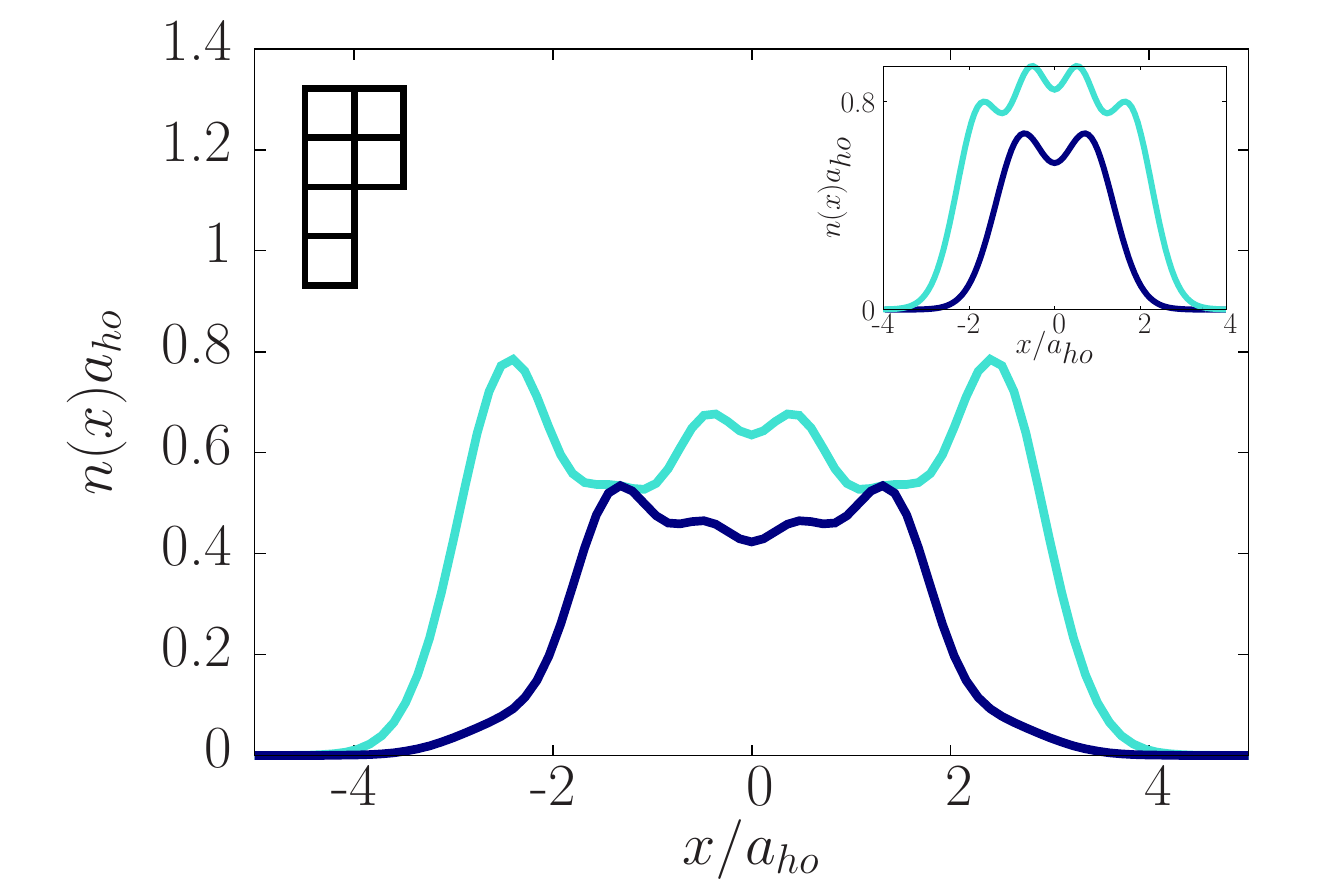}
\includegraphics[width=0.45\linewidth]{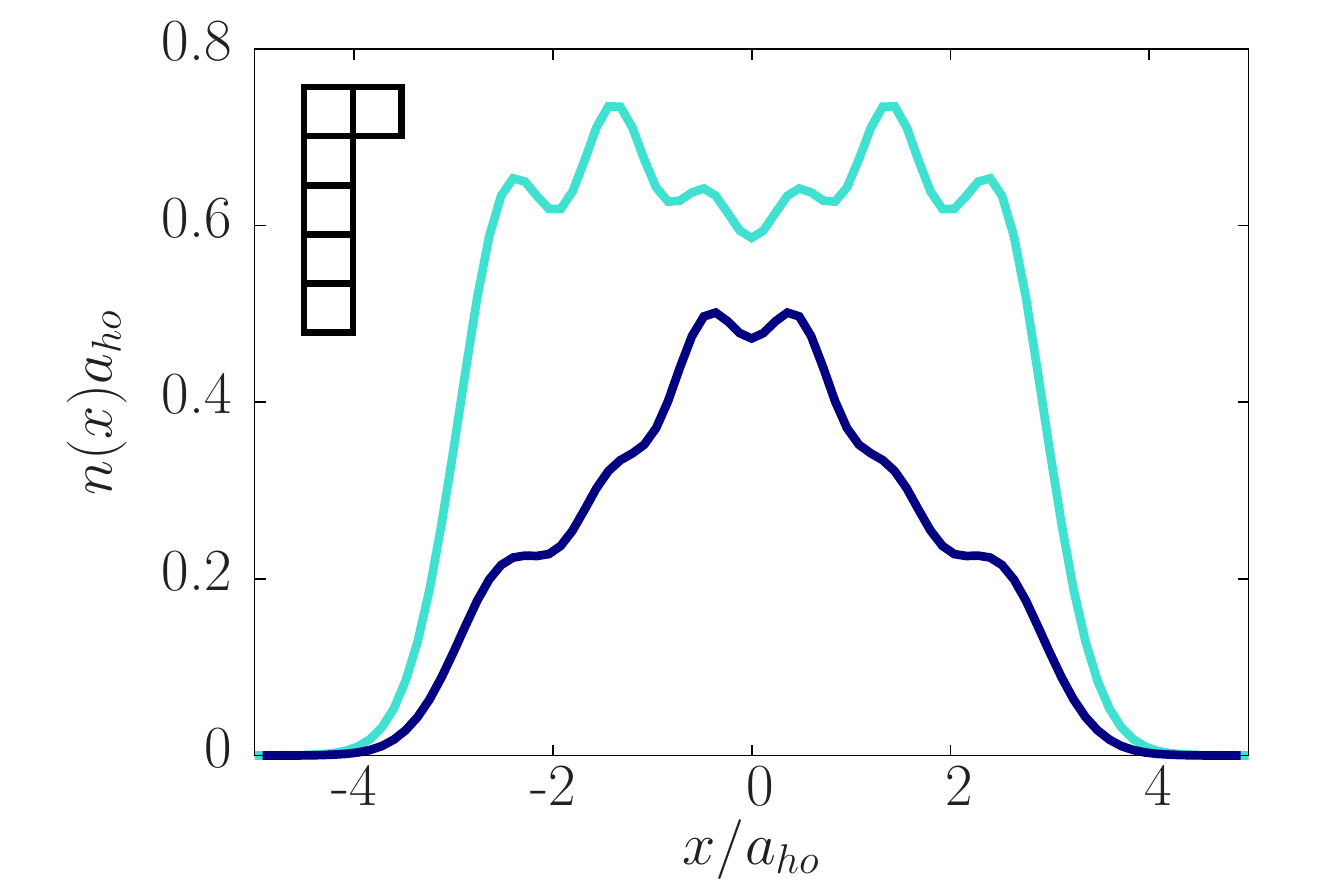}
\includegraphics[width=0.45\linewidth]{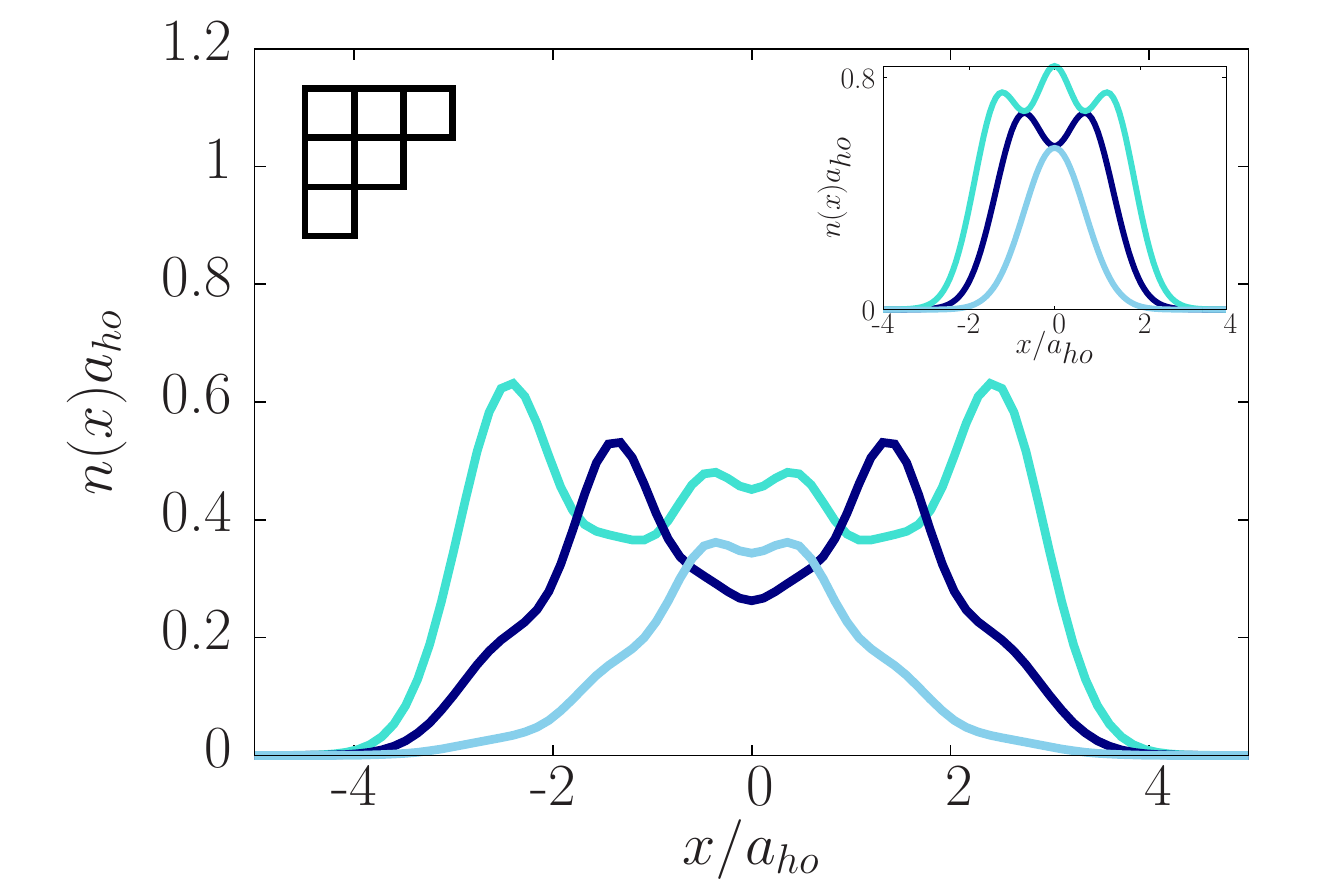}
\includegraphics[width=0.45\linewidth]{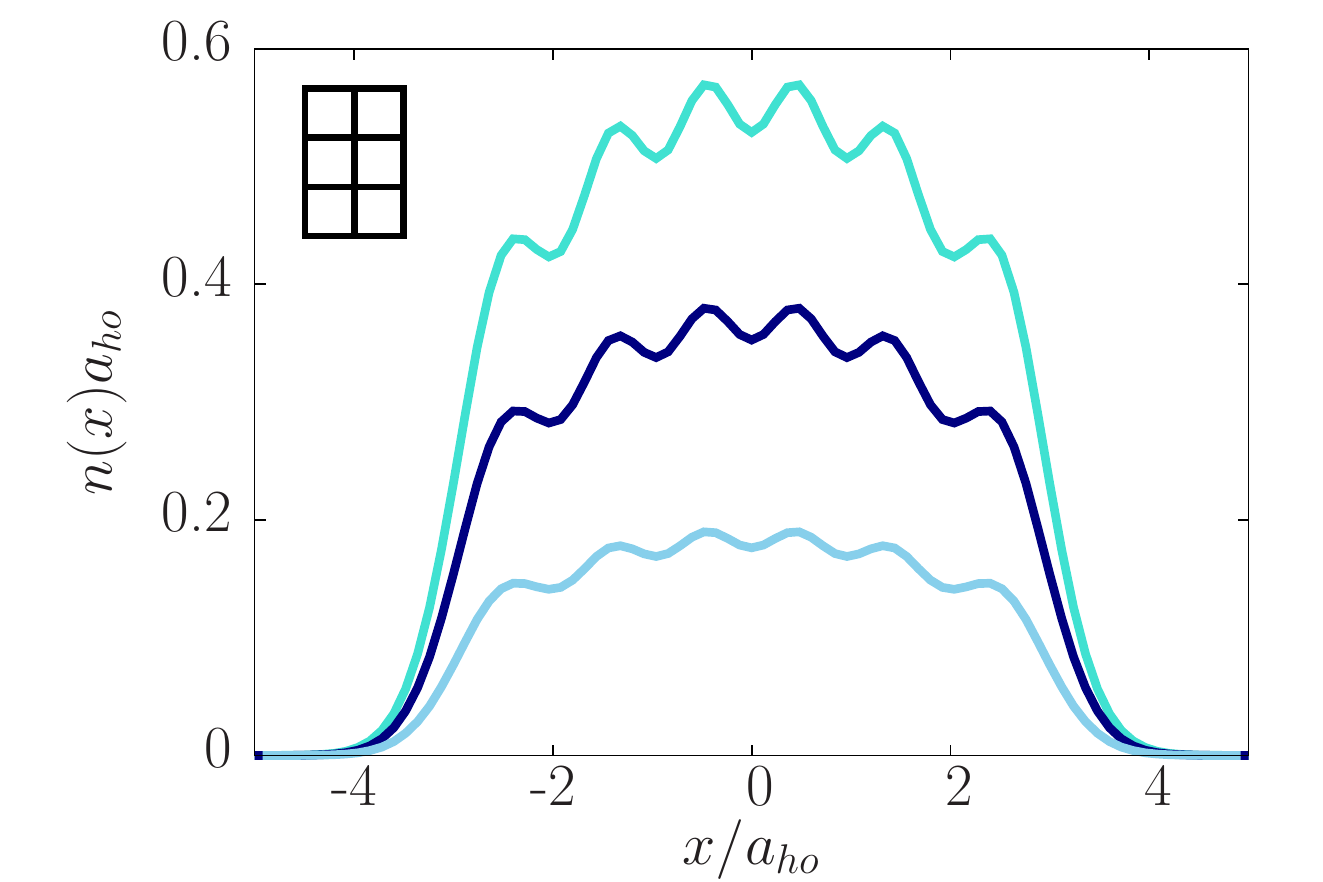}
\end{center}
\caption{\label{fig2} 
 Density profiles for the ground state (left panels) and for the first many-body excited state with a symmetry different than the ground state (right panels)
 for three imbalanced mixtures of strongly interacting Fermi gases, all having the same total number of particles $N=6$. Top panels: a  two-component mixture  with $N_1=5$ (turquoise) and $N_2=1$ (dark blue). Central panels: a two-component mixture  with $N_1=4$ (turquoise) and $N_2=2$ (dark blue). Bottom panels: a three-component mixture with $N_1=3$ (turquoise), $N_2=2$ (dark blue), $N_3=1$  (cyan). The insets show the corresponding ground-state density profiles for the same mixture of noninteracting gases. The corresponding Young tableau (see text) is is also shown in each panel.}
\end{figure}

In the case of an imbalanced mixture, the partial density profiles display a rich structure, as presented in Fig.\ref{fig2}.
We first consider the case of an imbalanced two-component mixture. In the polaron case, where $N_1=N-1$ and $N_2=1$ (top panels in Fig.\ref{fig2}),  for the ground-state density we observe a spatial separation of the polaron density profile at the center of the trap and the majority component at the wings of the trap. By comparing with the results for the corresponding noninteracting gas, one clearly sees the effect of repulsive interactions: the mutual repulsion among the two components push the majority component to a larger region of the trap and a hole is created around the polaron. For the excited state, we find that the density profiles are proportional to the ones of a noninteracting gas. We understand this as being related to the symmetry properties of this particular excited state -- as it will be discussed in detail in Sec.~\ref{sec:sym} below, we find that it has the same symmetry as the noninteracting gas--. 

In the case of more than one particle in the minority component (central panels  in Fig.\ref{fig2}) for the ground state density profiles, we observe a partial demixing through a more complex structure, with the majority component occupying both the inner core and the external wings of the density profiles. This may be viewed as a mesoscopic realization of an antiferromagnetic configuration \cite{Murmann2015}. The excited state density profiles have instead a two-peak (majority component) or one-peak (minority component) structure which recalls the ground state of the polaron. Also in this case, the analysis of the symmetry of the profiles brings an explanation, since we find that this excited state has the same symmetry as the ground state of the polaron.  

The three component imbalanced mixture (bottom panels  in Fig.\ref{fig2}) displays an even more complex spatial-separated shell structure in the ground state profiles, generalizing the two component case:  the minority component occupies the inner of the trap,  and the majority component the outer shells, with the intermediate component placed spatially between the other two. For the excited state, we find again a mixed state whose density profiles are proportional to the ones of a noninteracting gas. As it will be discussed in detail in Sec.~\ref{sec:sym} below, this is in agreement with the fact that this state has the same symmetry as the ground-state of the two-component balanced mixture.

\section{Symmetry characterization}
\label{sec:sym}

We now analyze the symmetry properties of the quantum many body states under exchange of particles \cite{Harshman2014,Harshman2015,Harshman2015b,Guan2014}. Obviously, single component free fermions are described by a totally anti-symmetric wavefunction. In the presence of  several components interacting among each other, the wavefunction must be totally antisymmetric under exchange of particles belonging to each component but the total wavefunction has a more involved symmetry under exchange of an arbitrary pair of particles.  The actual symmetry of the ground state and the excited states can be deduced from the properties of the permutation group of $N$ elements, $S_N$ \cite{Hamermesh}. More precisely, one can expand the eigenstates of the quantum system over the different irreducible representations of this group. We follow this route below, and  demonstrate that the ground state (and any excited state) has a well defined symmetry, ie can be described by a single irreducible representation, identified by a Young tableau.

The previous diagonalization process allows us to obtain a set of $S$ values of $K$, $K_1\le\dots\le K_S$, with $K_1=0$ and $K_S=K_{max}$, and a set of associated eigenvectors $\vec A_1,\dots, \vec A_S$ that correspond to decreasing-energy solutions at finite interactions. In order to completely characterize the symmetry of the various states $(K_\ell,\vec A_\ell)$ of the degenerate manifold at $g=\infty$, we determine to which irreducible representation of $S_N$, ie to which Young tableau, the solution $\vec A_\ell$ corresponds. The Young tableaux are defined in the standard fashion \cite{Hamermesh}:  elements belonging to the same line (column) are symmetric (antisymmetric) under exchange. Thus, for example, for 6 particles divided in 6 components, a completely antisymmetric state $\vec A_\ell$  will correspond to {\tiny\Yvcentermath1$\yng(1,1,1,1,1,1)$} whereas a completely symmetric state will correspond to {\tiny\Yvcentermath1$\yng(6)$}. Note that these two states are associated with a single-component non interacting Fermi gas and a single component Tonks gas, respectively.

In order to classify those states according to their symmetry, we use the  $k$-cycle class sums operators $\Gamma^{(k)}$ \cite{Katriel1993,Novolesky1994}, defined by $\Gamma^{(k)}=\sum_{i_1<...<i_k}(i_1...i_k)$, where $(i_1...i_k)$ is the cyclic permutation of particles $i_1,\ldots,i_k$ . In our system, on the basis of the coordinate sectors, $\Gamma^{(k)}$ is a $N!\times N!$ operator whose elements are $\Gamma^{(k)}_{ij}=(-1)^{(k-1)}$ if going from sector $i$ to sector $j$ only exchange $k$ fermions in a cyclic fashion, and $\Gamma^{(k)}_{ij}=0$ otherwise. Note that, also in this case, we can reduce the dimension of such operator to an $S\times S$ matrix, by summing over the contributions of the sectors associated with one snippet.

The spectral decomposition of $\Gamma^{(k)}$ allows to associate each Young tableau $Y$ with a given eigenvalue $\gamma_k$ \cite{Katriel1993,Novolesky1994,Fang2011}. In the following, we shall use in particular the class-sum operator $\Gamma^{(2)}$.   Our method consists first in computing and diagonalizing the transposition class sum $\Gamma^{(2)}$ for a given system. Its eigenvalues $\gamma_2$ can be linked to the  Young tableaux according to the expression
\begin{equation}
\gamma_2=\dfrac{1}{2}\sum_i\lambda_i(\lambda_i-2i+1),
\label{young}
\end{equation}
where $i$ and $\lambda_i$ refer respectively to the line and number of boxes in this line of Young tableau.
Thus, projecting a given solution $\vec{A}_\ell$ over the eigenbasis of $\Gamma^{(2)}$ allows to characterize its symmetry and to analyze it in terms of Young tableaux.
In Table 1 we summarize some of our results for the ground states $(\vec{A}_{\footnotesize{\mbox{max}}},K_{\footnotesize{\mbox{max}}})$ and the first excited states with a different symmetry $(\vec{A}_{\footnotesize{\mbox{excited}}},K_{\footnotesize{\mbox{excited}}})$ of different systems for $N=6$ particles.
\begin{table}
\begin{center}
\begin{tabular}{c}
\begin{tabular}{|c||c|c|c|c|}
\hline System & $A_{\footnotesize{\mbox{max}}}$ & $K_{\footnotesize{\mbox{max}}}$ & $A_{\footnotesize{\mbox{excited}}}$ & $K_{\footnotesize{\mbox{excited}}}$ \\
\hline $r=2$, $N_1=N_2=3$ & $Y_{-3}$  & $24.97$ & $Y_{-5}$ & $18.91$ \\
\hline $r=3$, $N_1=N_2=N_3=2$ & $Y_3$ & $30.63$ & $Y_0$ & $28.96$ \\
\hline $r=6$,  $N_1=...=N_6=1$ & $Y_{15}$ & $34.33$ & $Y_9$ & $33.35$ \\
\hline $r=2$, $N_1=5$, $N_2=1$ & $Y_{-9}$  & $14.60$ & $Y_{-15}$ & $0$\\
\hline $r=2$, $N_1=4$, $N_2=2$ & $Y_{-5}$   &  $22.70$ & $Y_{-9}$ & $14.60$ \\
\hline $r=3$, $N_1=3$, $N_2=2$, $N_3=1$ & $Y_{0}$ & $28.96$ & $Y_{-3}$ & $24.97$\\
\hline
\end{tabular}
\end{tabular}
\end{center}
\label{tableau}
\caption{Symmetries and corresponding $K$ eigenvalue (see Eq.(\ref{Eq:inlineVAeqKA})) for the multi-component fermionic mixtures of Figs.~\ref{fig1} and \ref{fig2}.}
\end{table}
We set \cite{Hamermesh}
\begin{equation}
\begin{split}
&
\begin{tabular}{cccccccc}
$Y_{15}$ = &\Yvcentermath1$\yng(6)$~,
 &~& $Y_{-9}$ = &\Yvcentermath1 $\yng(2,1,1,1,1)$~,
 &~& $Y_{-5}$ = &\Yvcentermath1$\yng(2,2,1,1)$~,
\end{tabular}\\&
\begin{tabular}{cccccccc}
 $Y_{-3}$ = &\Yvcentermath1$\yng(2,2,2)$~,
 &~& $Y_{0}$ = &\Yvcentermath1$\yng(3,2,1)$~,
\end{tabular}
\end{split}
\end{equation}
and the $Y_{-\gamma}$ are obtained by taking the symmetric of  $Y_{\gamma}$ with respect to  the main diagonal.

Table 1 shows that the ground and excited states constructed with Eq.(\ref{Eq3:Psi}) have a definite symmetry which can be readily extracted from the associated  Young tableau. An exception is provided by the ground state of the case $N_1=N_2=3$ as well as the excited state of $N_1=3$, $N_2=2$, $N_3=1$, where the transposition class-sum operator $\Gamma^{(2)}$ does not allow to uniquely associate a Young tableau to the wavefunction, since  Eq.~\eqref{young} gives two Young tableaux corresponding to   the eigenvalue $-3$, {\tiny\Yvcentermath1$\yng(2,2,2)$} and {\tiny\Yvcentermath1$\yng(3,1,1,1)$}, and also to the eigenvalue $3$, {\tiny\Yvcentermath1$\yng(3,3)$} and {\tiny\Yvcentermath1$\yng(4,1,1)$}.  In this case, the ambiguity is lift off with the help of the 3-cycle class sum $\Gamma^{(3)}$, since these two Young tableaux correspond to two different eigenvalues of $\Gamma^{(3)}$ ~\cite{Katriel1993}. 

Furthermore, for all the mixture considered we note  that  the ground state, corresponding to $K=K_{max}$, is the most symmetric one \cite{Deuretzbacher15, Deuretzbacher2016}, and the excited state is obtained by decreasing the symmetry of the state by taking out one cell from the top row and putting it in the first available lower row of the Young tableau, thus making it more antisymmetric. The opposite case of $K=0$ is associated with the most antisymmetric Young tableau.  The above result for the ground state supports the observation of Ref.~\cite{MassignanPaarish}, where an ansatz for the ground state wavefunction was suggested, and is also in agreement with a general demonstration provided in \cite{Hakobian10}.  We notice also that  when the number of particles coincides with the number of components, ie $r=N$,  the ground state is fully symmetric, and has the same symmetry as the bosonic Tonks-Girardeau gas.

This analysis provides a verification and an example of the  generalization of the Lieb-Mattis theorem \cite{LiebMattisPR} for the case of multicomponent fermionic mixtures. The theorem states that for $N$ electrons in one dimension,  interacting by a symmetric potential,  the energy of a state with total spin $S$ $(S')$ is such that $E(S)\le E(S')$ if $S<S'$. It follows that the ground state has the smallest possible value for the spin $S$, which is realized by an antisymmetric spinor. Hence, the spatial wavefunction has the most symmetric configuration. In the case of more than two spin components, as also discussed in \cite{Deuretzbacher15}, the same feature occurs, which is displayed by our results. Since $K$ tends to be maximized when the wavefunction is more symmetric, we can see $K$ as an energetic indicator of the symmetry.

\section{Summary and conclusions}

In this work we have considered a multicomponent strongly correlated fermionic mixture with up to six components. Using a generalization of the pioneering solution due to Girardeau for the Bose gas and of the recently developed solution for the two-component Fermi gas we have determined the exact many-body wavefunctions for the degenerate manifold at infinite interactions. We have identified the one which corresponds to the ground state at finite interactions as the one which has the maximum value for the parameter $K$, related to the interaction energy. We have then obtained the density profiles for the mixture under harmonic confinement. For an imbalanced mixture we have found a partial phase separation among the components, which is an effect of the strong repulsive interactions.
Furthermore, we have characterized the symmetry properties of the ground and of some excited state wavefunctions of the manifold by introducing suitable class-sum operators, and we have shown that the ground-state wavefunctions have a definite symmetry, corresponding to the most symmetric (or less antisymmetric) configuration compatible with the imbalance among the components.
Our exact solution for the inhomogeneous multicomponent mixture provides an important benchmark for numerical simulations of strongly correlated multicomponent Fermi gases, in a regime where the presence of the quasi-degenerate manifold challenges the convergence of the calculations, as well as for quantum simulators.

\section*{Acknowledgements}
This paper is dedicated to the memory of Marvin Girardeau, of whom AM recalls  his enthusiastic guidance. We thank M. Gattobigio, J. Levinsen, P. Massignan, M. Parish and T. Roscilde  for fruitful discussions. AM acknowledges the hospitality of the Centre de Recherches Math\'ematiques de Montr\'eal where part of this work was performed  and  financial support from ANR projects Mathostaq (ANR-13-JS01-0005-01) and  SuperRing (ANR-15-CE30-0012-02).
 
\def\newblock{\hskip .11em plus .33em minus .07em} 


\end{document}